\documentclass[aps,floatfix,onecolumn,a4paper,nofootinbib,showpacs, 11pt,superscriptaddress]{revtex4}
\usepackage{graphicx,float}
\usepackage[all]{xy}
\usepackage{amsmath}
\usepackage{amssymb}
\usepackage{color}
\usepackage{epsfig}		

\newcommand{\bes}{\begin{subequations}}
\newcommand{\ees}{\end{subequations}}
\def\ben{\begin{eqnarray}}
\def\een{\end{eqnarray}}
\def\be{\begin{equation}}
\def\ee{\end{equation}}

\begin{document}
\title{Matter localization on brane-worlds generated by deformed defects} 
\author{Alex E. Bernardini} 
\affiliation{Departamento de F\'isica, Universidade Federal de S\~ao Carlos,
PO Box 676, 13565-905, S\~ao Carlos, SP, Brasil}
\author{Rold\~ao da Rocha}
\affiliation{Centro de Matem\'atica, Computa\c c\~ao e Cogni\c c\~ao, Universidade Federal do ABC, UFABC, 09210-580, Santo Andr\'e, Brazil}
\vspace{1 cm}

\begin{abstract}
\begin{center}
{\bf Abstract}
\end{center}
Localization and mass spectrum of bosonic and fermionic matter fields of some novel families of asymmetric thick brane configurations generated by deformed defects are investigated.
The localization profiles of spin 0, spin 1/2 and spin 1 bulk fields are identified for novel matter field potentials supported by thick branes with internal structures. The condition for localization is constrained by the brane thickness of each model such that thickest branes strongly induces matter localization. 
The bulk mass terms for both fermion and boson fields are included in the global action as to produce some imprints on mass-independent potentials of the Kaluza-Klein modes associated to the corresponding Schr\"odinger equations.
In particular, for spin 1/2 fermions, a complete analytical profile of localization is obtained for the four classes of superpotentials here discussed.
Regarding the localization of fermion fields, our overall conclusion indicates that thick branes produce a {\em left-right asymmetric chiral} localization of spin 1/2 particles. 
\end{abstract}
\pacs{11.25.Mj, 04.40.Nr, 11.10.Kk}
\maketitle
\date{\today}
 

\section{Introduction}
 
 The brane-world model is a prominent paradigm that has been addressed to solve several questions in physics. Within this framework, brane-worlds are required to render a consistent 4D physics of our Universe, at least up to certain sensible limits \cite{Arkani02}. In the brane-world scenario all kinds of matter fields should be localized on the brane. In the RS brane-world model \cite{Randall}, the brane is generated by a scalar field coupled to gravity \cite{DeWolfe,Gremm}, in a particular scenario which may be interpreted as the thin brane limit of thick brane scenarios.
Generically, a prominent test that thick brane-world models must pass, to be physically consistent, regards their stability, with respect to tensor, vector, and scalar fluctuations of the background fields that generate the field configurations, namely, the thick brane itself. 
At least the zero modes of Standard Model matter fields were shown
to be localized on several brane-world models \cite{Liu0907.0910, Liu_20101,Zhang,Andrianov:2013vqa,Folomeev,Brane}, suggesting that such kind of models are physically viable in high energy physics. Several alternative scenarios, including Gauss-Bonnet terms, $f(R)$ gravity, tachyonic potentials, cyclic defects, and Bloch branes, have been further studied \cite{German:2012rv,German:2013sk,Bernardini,ddutra}, and analogous scenarios in an expanding Universe have been approached \cite{Ahmed,Bernardini:2014vba}. {The curvature nature  of the brane-world, namely, to be a de Sitter, Minkowski or anti-de Sitter one, is in general  obtained \emph{a posteriori}, by solving the 5D Einstein field equations. In fact,  
the bulk and the brane cosmological constants depend upon the brane and the bulk gravitational  field content, governed by curvature, and must obey  the intrinsic fine-tuning, in the Randall-Sundrum-like models limit.}

The analytical study of stability can be uncontrollably intricate, due to the involved
structure of the scalar field coupled to gravity. To circumvent the complicated and not analytical approaches, linearized formulations have been commonly worked out. In this context, supported by the stability of deformed defect generated brane-world models, scalar, vector, and tensor perturbations are investigated throughout this work.
 
Localization aspects of various matter fields with spin 0, 1/2, and 1 on analytical thick brane-world models are indeed a main concern in deriving brane-world models, since they must describe our physical $4D$ world. The localization of the spin 1/2 fermions deserves a special attention, since there is no scalar field to couple with in this model, in contrast to thick branes generated by deforming defect mechanisms \cite{Barbosa-Cendejas:2015qaa}. 
Otherwise, Kalb-Ramond fields, although already investigated \cite{Cruz:2013zka}, shall not be the main aim here. 
The spin 1/2 issue has been previously studied in some other contexts \cite{PRD_Oda_026002}, including further coupling of more scalar fields in the action \cite{casana} and asymmetric brane-worlds generated by a plenty of scalar field potentials \cite{bazeiamr,Bazeia:2012qh,Guo:2011qt,Dutra:2013jea,Liu0907.0910}. In particular, asymmetric Bloch branes in the context of the hierarchy problem have been addressed in Ref. \cite{ddutra}. 

Our aim is to investigate the localization of bulk matter and gauge fields on the brane, in the context where the mass-independent potentials of the corresponding Schr\"odinger-like equations, regarding the 1$D$ quantum mechanical analogue problem, can be suitably acquired from a warped metric. In particular, for a bulk mass proportional to the fermion mass term enclosed by the global action, the possibility of trapping spin 1/2 fermions on asymmetric branes is discussed and quantified. 

To accomplish this aim, this paper is organized as follows.
In Sect. II, a brief review of brane-world scenarios supported by an effective action driven by a (dark sector) scalar field is presented. Warp factors and the corresponding internal brane structure are described for four different analytical models.
In Sect. III, the {\em left-right} chiral asymmetric aspects of matter localization for spin 1/2 fermion fields on thick branes are investigated.
Extensions to scalar boson and vector boson fields are obtained in Sect. IV and V, respectively.
Final conclusions are drawn in Sect. VI.

\section{Brane-world preliminaries and some analytical models}

Let one starts considering a $5D$ space-time warped into $4D$.
The most general $5D$ metric compatible with a brane-world spatially flat cosmological background has the form given by
\begin{eqnarray}
 ds^2 &=& g_{MN}dx^{M}dx^{N}
     ={\rm e}^{2A(y)}\,{\rm g}_{\mu\nu}(x^\alpha)dx^\mu dx^\nu + dy^{2}\label{metricaa}
    \end{eqnarray}
where $\text{e}^{2A(y)}$ denotes the warp factor, and the signature $(-++++)$ is employed, with $M,N=0,\,1,\,2,\,3,\,5$. The ${\rm g}_{\mu\nu}$ stands for the components of the $4D$ metric tensor ($\mu,\nu=0,1,2,3$).
One can identify $y \equiv x_4$ as the infinite extra-dimension coordinate (which runs from $-\infty$ to $\infty$), and notice that the normal to surfaces of constant $y$ is orthogonal to the brane, into the bulk \footnote{Brane tension terms have been suppressed/absorbed by the metric (c. f. Eqs. (24) and (25) from Ref. \cite{Folomeev} for real scalar field Lagrangians in the context of thick brane solutions).}.

The brane-world scenario examined here is setup by an effective action, driven by a (dark sector) scalar field, $\zeta$, coupled to $5D$ gravity, given by
\begin{equation}
S_{\tiny\mbox{eff}} = - \int dx^5\, \sqrt{\det{g_{MN}}}\,\left[\frac{1}{4}({\kappa_5^{-2}}R{-2\Lambda_5}) + \frac{1}{2}g_{MN}\partial^M \zeta\partial^N \zeta - V(\zeta)\right],
\label{eqs000}
\end{equation}
where $R$ is the $5D$ scalar curvature, $R_{NQ}=g^{BM}R_{BNQM}$ is the Ricci tensor, {and  $\kappa_5 = (8\pi G_5)^{1/2}$ denotes the 5D gravitational coupling constant, hereon set to equal to unity, where $G_5$ is the 5D Newton constant.}
The Einstein equations read \begin{eqnarray}
R_{MN}-\frac{1}{2}R\ g_{MN}=-\Lambda_{5}\ g_{MN}{+\kappa_5^2 T_{MN}},
\end{eqnarray} {where $T_{MN}^\zeta$
 denotes the energy-momentum tensor corresponding to 
 the matter Lagrangian, regarding the matter field $\zeta$}. 
{After solving the 5D Einstein field equations, 
the bulk cosmological constant turns out, in general, to be
positive or negative, thus realising a de Sitter or anti-de Sitter brane-world, respectively, 
generated by curvature. It realises and emulates the interplay involving the
4D and 5D cosmological constants. Some further possibilities are devised, e. g., in \cite{German:2012rv,HerreraAguilar:2010kt}, however it is worth to mention that an additional scalar field can be still added in the action, whose isotropisation shall precisely define the nature of the brane-world. This the latter case is however beyond the scope of our analysis. Obviously, whatever the possibility to be considered, the thin brane limit must obey the fine-tuning relation ~\cite{maartens}
$
\Lambda_4=\frac{\kappa_5^{2}}{2}\left(\frac{1}{6}\kappa_5^{2}\sigma^{2}+\Lambda_{5}\right)
$, among the effective 4D and 5D cosmological constants and the brane tension
$\sigma$ as well. }

Considering the real scalar field action, Eq.~(\ref{eqs000}), one can compute the stress-energy tensor
\begin{equation}
T_{MN}^{\zeta} = \partial_M \zeta\partial_N \zeta + g_{MN}\,V(\zeta) - \frac{1}{2}g_{MN}\,g^{AB} \partial_A \zeta \partial_B \zeta,
\label{eqs002}
\end{equation}
which, supposing that both the scalar field and the warp factor dynamics depend only upon the extra coordinate, $y$, leads to an explicit dependence of the energy density in terms of the field, $\zeta$, and of its first derivative, $d\zeta/dy$, as
\begin{equation}
T_{00}^{\zeta} (y) =\left[\frac{1}{2}\left(\frac{d\zeta}{dy}\right)^2 + V(\zeta)\right]\, {\rm e}^{2A(y)}.
\label{eqs003}
\end{equation}

With the same constraints on $\zeta$ about the dependence on $y$, the equations of motion currently known from \cite{DeWolfe,Gremm}, which arise from the above action, are
\begin{equation}
\frac{d^2\zeta}{dy^2} + 4 \frac{d A}{dy} \frac{d \zeta}{dy} - \frac{d}{d\zeta}V(\zeta) = 0,
\label{eqs004}
\end{equation}
through a variational principle relative to the scalar field, $\zeta$, and
\begin{equation}
\frac{3}{2}\frac{d^2 A}{dy^2} = - \left(\frac{d\zeta}{dy}\right)^2,
\label{eqs005}
\end{equation}
through a variational principle relative to the metric, or equivalently to $A$, manipulated to result into 
\begin{equation}
3 \left(\frac{d A}{dy}\right)^2 = \frac{1}{2} \left(\frac{d\zeta}{dy}\right)^2 - V(\zeta), 
\label{eqs006}
\end{equation}
after an integration over $y$.

For the scalar field potential written in terms of a {\em superpotential}, $w$, as 
\begin{equation}
V(\zeta) = \frac{1}{8}\left(\frac{dw}{d\zeta}\right)^2 - \frac{1}{3} w^2,
\label{eqs007}
\end{equation}
the above equations are mapped into first-order equations \cite{DeWolfe,Gremm} as
\begin{equation}
\frac{d\zeta}{dy} = \frac{1}{2}\frac{d w}{d\zeta},
\label{eqs008}
\end{equation}
and
\begin{equation}
\frac{d A}{dy} = - \frac{1}{3} w,
\label{eqs009}
\end{equation}
for which the solutions can be found straightforwardly through immediate integrations \cite{DeWolfe} (see also Ref. \cite{Folomeev} and references therein).
The energy density follows from Eq.~(\ref{eqs007}) as
\begin{equation}
T_{00}^{\zeta}(y) = \left[\frac{1}{4}\left(\frac{dw}{d\zeta}\right)^2 - \frac{1}{3} w^2\right]\, {\rm e}^{2A(y)}\label{eqs0031}.
\end{equation}

The analysis of localization aspects of brane-world scenarios shall be constrained by some known examples, $I$, $II$, $III$ and $IV$, for which the warp factor, $A(y)$, and the energy density, $T_{00}(y)$, can be analytically computed.
The model $I$ is supported by a sine-Gordon-like superpotential given by
\begin{equation}
w^I(\zeta) = \frac{2}{\sqrt{2} a} \sin{\left(\sqrt{\frac{2}{3}} \zeta\right)},
\label{014}
\end{equation}
which reproduces the results from Ref. \cite{Gremm}.
The model $II$ corresponds to a deformed $\lambda \zeta^4$ theory with the superpotential given by
\begin{equation}
w^{II}(\zeta) = \frac{3\sqrt{3}}{a} \left(1 - \frac{\zeta^2}{9}\right)^{3/2}.
\label{0141}
\end{equation}
Models $III$ and $IV$ are deformed topological solutions from Ref. \cite{Bernardini13} supported by superpotentials like 
\begin{equation}\label{sp001}
w^{III}(\zeta) = \frac{2}{a} \arctan\left[\sinh(\zeta)\right],
\end{equation}
and 
\begin{equation}\label{sp002}
w^{IV}(\zeta) = \frac{1}{4a}\left[\zeta\left(5 - 2 \zeta^2\right)\sqrt{1 - \zeta^2} 
+ 3 \arctan\left(\frac{\zeta}{\sqrt{1 - \zeta^2} }\right)\right],
\end{equation}
where the parameter $a$ fixes the thickness of the brane described by the warp factor, ${\rm e}^{2A(y)}$.
Besides exhibiting analytically manipulable profiles, the above superpotentials have already been discussed in the context of thick brane localization \cite{Gremm,Brane,Folomeev}.
The models $I$ and $II$ are respectively motivated by sine-Gordon and $\lambda \zeta^{4}$ theories, and models $III$ and $IV$ are obtained (also analytically) from deformed versions of the $\lambda \zeta^{4}$ model \cite{Bernardini}.
In particular, models $III$ and $IV$ can also be mapped onto tachyonic Lagrangian versions of scalar field brane models
\cite{Bernardini13,Zhang,Folomeev}.

From the above superpotentials, the respective solutions for $\zeta(y)$ are set as 
\begin{eqnarray}\label{015}
\zeta^{I}(y) &=& \sqrt{6} \arctan{\left[\tanh\left(\frac{y}{2\sqrt{2}a}\right)\right]},\\
\zeta^{II}(y)&=& 3 \, \mbox{sech}{\left(\frac{\sqrt{3}y}{2a}\right)},\\
\zeta^{III}(y) &=& \mbox{arcsinh}\left(\frac{y}{a}\right),\\
\zeta^{IV}(y) &=& \frac{y}{\sqrt{a^2 + y^2}}\label{015B}
,
\end{eqnarray}
where one has suppressed any additional (irrelevant) constant of integration
 for convenience, and one has just considered the positive solutions \footnote{In Eqs.~(\ref{015})-(\ref{015B}) there could let be explicit a constant of integration that amounts to letting $y \mapsto y + C$, corresponding to the position of the brane in the extra dimension, for which one has set $C = 0$.}.

The obtained expressions for the warp factor as resulting from Eq.~(\ref{eqs009}) are respectively given by
\begin{eqnarray}
A^{I}(y) &=& - \ln{\left[\cosh\left(\frac{y}{\sqrt{2}a}\right)\right]},\\
A^{II}(y) &=& \tanh{\left(\frac{\sqrt{3}y}{2a}\right)}^2 - 2 \ln{\left[\cosh\left(\frac{\sqrt{3}y}{2a}\right)\right]},\\
A^{III}(y) &=& \frac{1}{3}\left[\ln\left(1 + \frac{y^2}{a^2}\right) - 2\frac{y}{a} \arctan\left(\frac{y}{a}\right)\right],\\
A^{IV}(y) &=& -\frac{1}{12}\left[\frac{y^2}{a^2 + y^2} + 3\frac{y}{a} \arctan\left(\frac{y}{a}\right)\right].
\label{0161}
\end{eqnarray}
where integration constants are introduced as to set a normalization criterium for which $A(0) = 0$.

\begin{figure}
\centerline{\psfig{file= 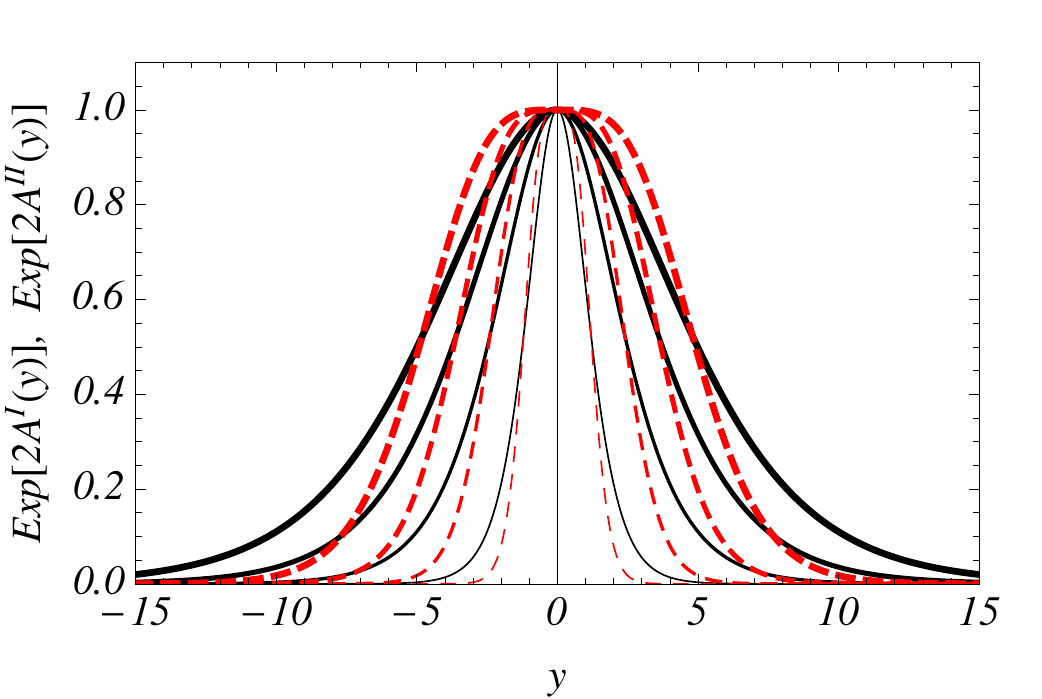, width= 10 cm}}
\centerline{\psfig{file= 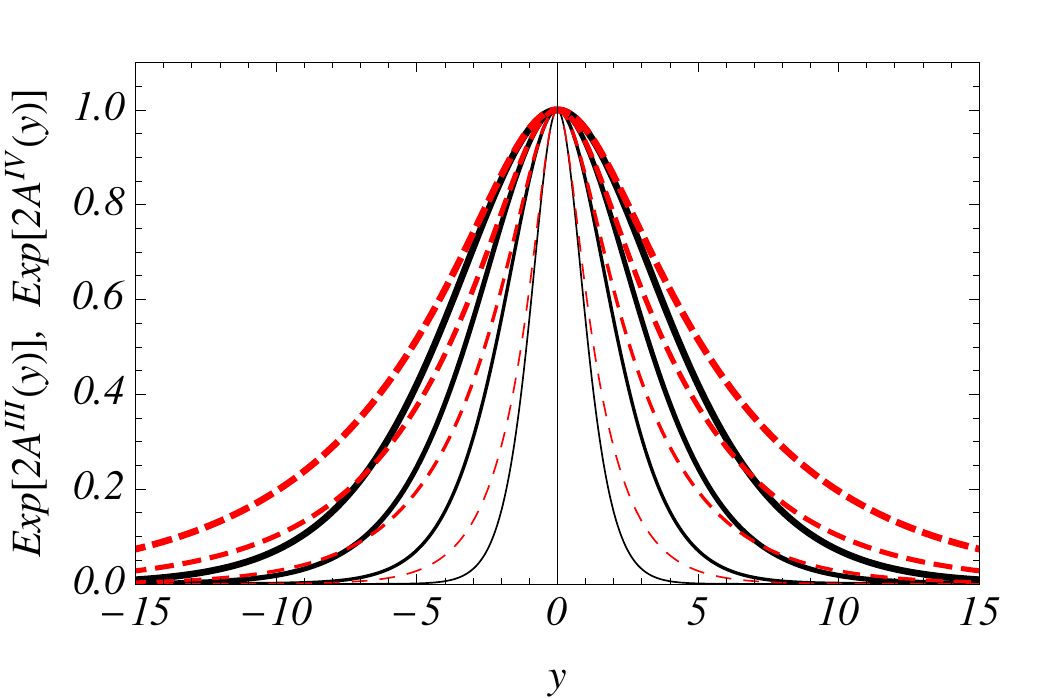, width= 10 cm}}
\caption{(Color online) {\em First plot:} warp factors, ${\rm e}^{2A(y)}$ for models $I$ (solid (black) lines) and $II$ (dashed (red) lines). {\em Second plot:} warp factors, ${\rm e}^{2A(y)}$ for models $III$ (solid (black) lines) and $IV$ (dashed (red) lines).
One has considered integer values of the brane width parameter $a$, running from $1$ (thinest line) to $4$ (thickest line), corresponding to an increasing thickness.}
\label{brane01}
\end{figure}

The solutions for $A^{I}$ and $A^{II}$ are depicted in Fig. \ref{brane01}.
The corresponding localized energy densities computed through Eq.~(\ref{eqs0031}) are respectively given by
\begin{eqnarray}
T^{I}_{00}(y) &=& \frac{3}{4a^2}\mbox{sech}\left(\frac{y}{\sqrt{2}a}\right)^2
\left[\mbox{sech}\left(\frac{y}{\sqrt{2}a}\right)^2 - 2 \tanh\left(\frac{y}{\sqrt{2}a}\right)^2\right],\\
T^{II}_{00}(y) &=&\frac{9}{8a^2}\, 
\mbox{sech}\left(\frac{\sqrt{3}y}{2a}\right)^{8}
\tanh\left(\frac{\sqrt{3}y}{2a}\right)^2 \left[
7\cosh\left(\frac{\sqrt{3}y}{a}\right)-\cosh\left(\frac{2\sqrt{3}y}{a}\right)\right]
{\rm e}^{2\tanh\left(\frac{\sqrt{3}y}{2a}\right)^2}
,\qquad\\
T^{III}_{00}(y) &=&\frac{1}{a^2} \left(1+\frac{y^2}{a^2}\right)^{-1/3} \left(1-\frac{4}{3}\left(1+\frac{a^2}{y^2}\right)\arctan\left(\frac{y}{a}\right)^{2}\right)
{\rm e}^{\frac{4y}{3a}\arctan\left(\frac{y}{a}\right)},
\\
T^{IV}_{00}(y) &=&\left(\frac{a^4}{(a^2+y^2)^3} -\frac{5 a^3 y+3 a y^3 + 3 (a^2+y^2)^2\arctan\left(\frac{y}{a}\right)^{2}}{48 a^2 (a^2+y^2)^4}\right)
{\rm e}^{-\frac{y}{2a}\left( \frac{a y}{3(a^2+y^2)}+\arctan\left(\frac{y}{a}\right)\right)}.
\label{0201}
\end{eqnarray}

\begin{figure}
\centerline{\psfig{file= 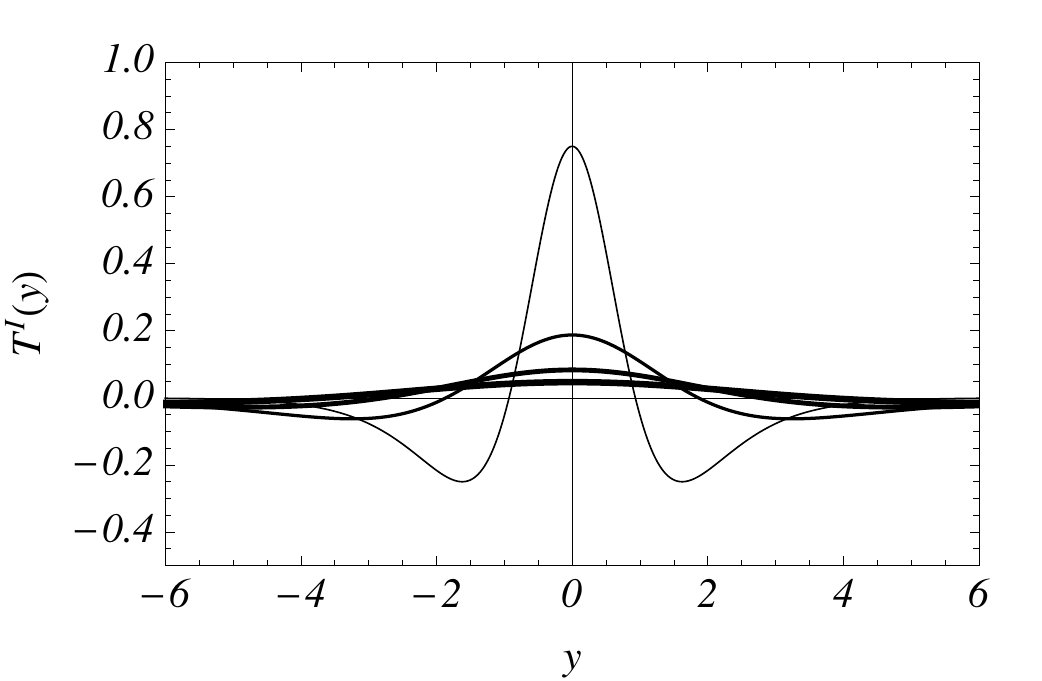, width= 8.5 cm}}
\centerline{\psfig{file= 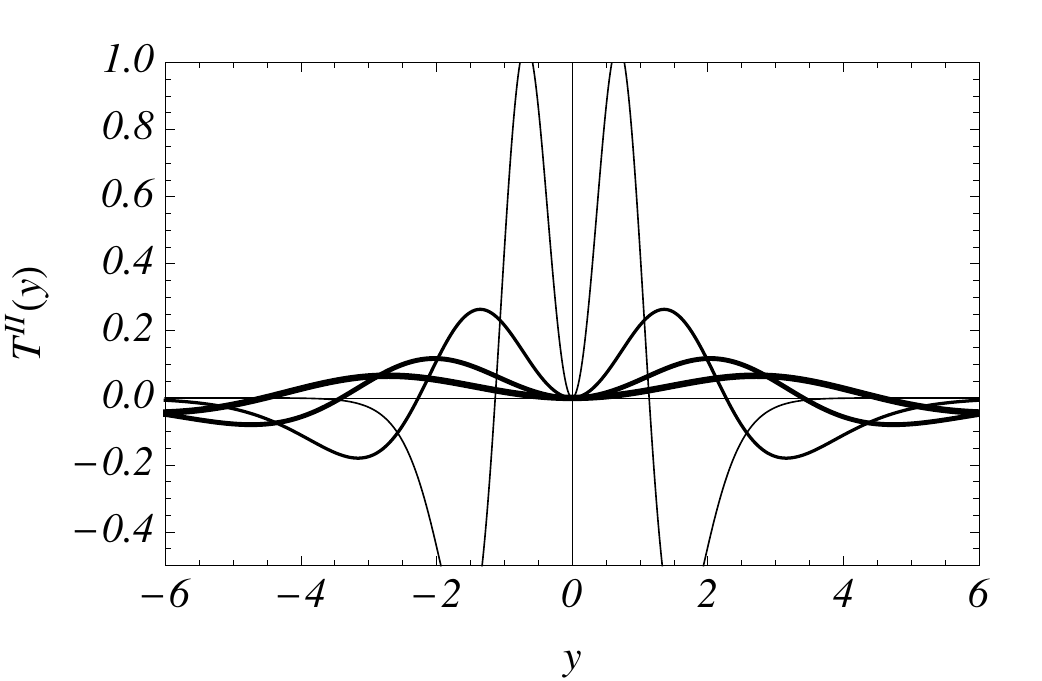, width= 8.5 cm}}
\centerline{\psfig{file= 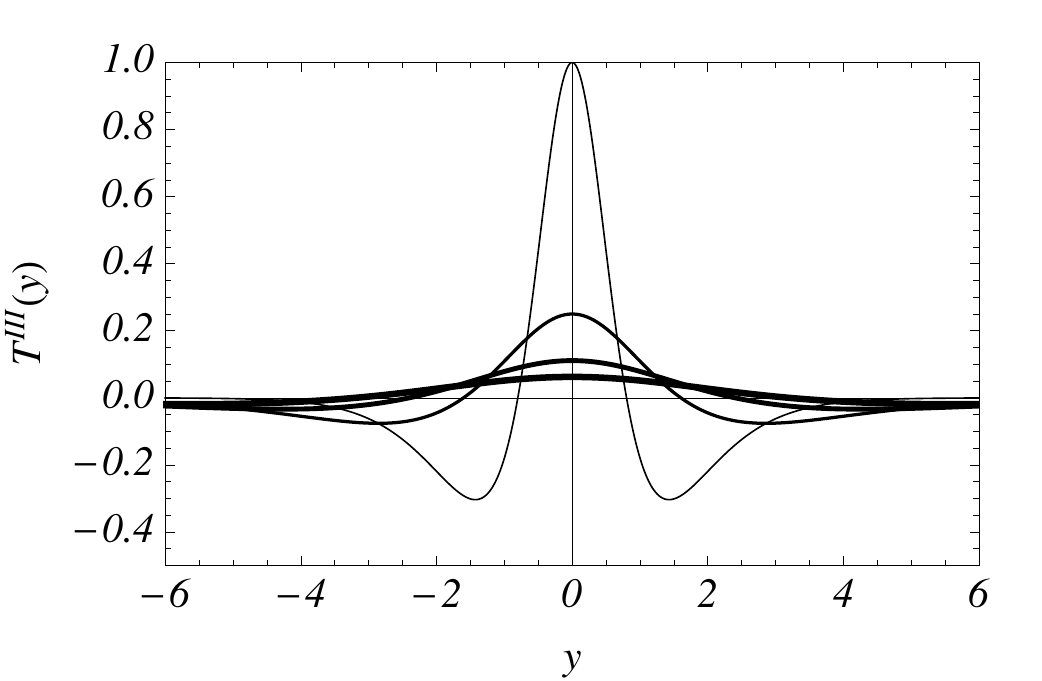, width= 8.5 cm}}
\centerline{\psfig{file= 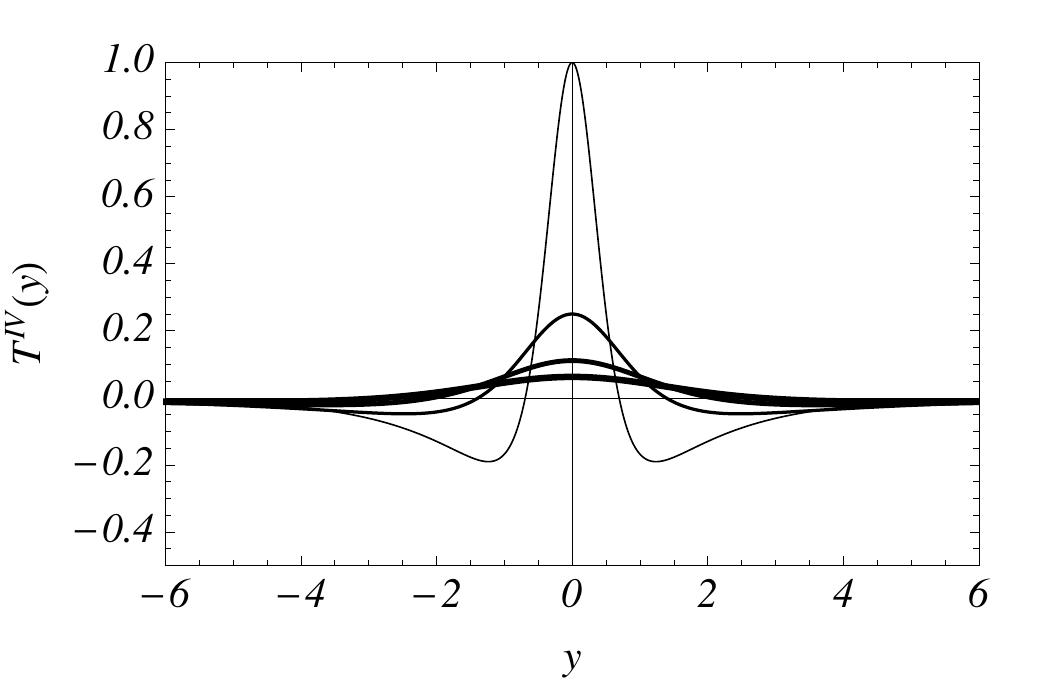, width= 8.5 cm}}
\caption{Energy density, $T_{00}(y)$, for models from $I$ (first plot) to $IV$ (fourth plot), where integer values of the brane width parameter $a$, running from $1$ (thinest line) to $4$ (thickest line), corresponding to an increasing thickness.}
\label{brane02}
\end{figure}

The brane scenarios for models from $I$ to $IV$ are depicted in Fig. (\ref{brane01}) for the warp factors and in Fig. (\ref{brane02}) for the energy densities, from which one can observe that models from $I$ to $IV$ give rise to thick branes, most of them with no internal structures.
In fact, only the potential that controls the scalar field from model $II$ allows the emergence of thick branes that host internal structures in the form of a layer of a novel phase enclosed by two separate interfaces, inside which the energy density of the matter field gets more concentrated. 
It is related to the extension/localization of the warp factor, namely when the profiles depicted in Fig. (\ref{brane01}) approach to a {\em plateu} form in the region very inside of the brane, the corresponding internal structure is observed through its energy profile. 

The appearance of negative energy densities in the plots for $T_{00}$ may be related to a predominance of the scalar field potential over the kinetic-like term related to the coordinate $y$.
Speculatively, it indicates that the vacuum minimal energy can be adjusted by the inclusion of some additional term, eventually related to the cosmological constant.

The localization of bulk matter fields on thick branes generated by each one of these models shall be identified in the following sections. Spin 0, spin 1/2, and spin 1 fields shall evolve coupled to gravity and, as usual, the bulk matter field contribution to the bulk energy shall be neglected. It means that the obtained solutions hold in the presence of the bulk matter, without disturbing the bulk geometry. 

\section{Asymmetric {\em left-right} matter localization for spin 1/2 fermion fields}

To investigate the localization of bulk matter on the brane, one first considers that fermion localization on brane-worlds is usually accomplished when the $5D$ Dirac algebra is realized by the objects $\Gamma^M= e^M_{~\bar{M}} \Gamma^{\bar{M}}$, where $e^M_{~\bar{M}}$ denotes the {\em f\"unfbein}, the $\Gamma^M$ satisfy the Clifford relation $\{\Gamma^M,\Gamma^N\}=2g^{MN}$, and $\Gamma^{\bar{M}}$ are the gamma matrices in the $5D$
flat spacetime.
Hereupon $\bar{M}, \bar{N}, \ldots =0,1,2,3,5$ and
$\bar{\mu}, \bar{\nu}, \ldots =0,1,2,3$ denote the $5D$ and $4D$ local
Lorentz indexes, respectively. The {\em f\"unfbein} is provided by
$
e_M ^{~\bar{M}}= \{\text{e}^{A} \tilde{e}_\mu^{~\bar{\nu}}, \text{e}^{A}\}$, where  
$\Gamma^M=\text{e}^{-A}(\gamma^{\mu},\gamma^5)$, and 
 $\gamma^{\mu}=\tilde{e}^{\mu}_{~\bar{\nu}}\gamma^{\bar{\nu}}$ and $\gamma^5$ are respectively the $4D$ gamma
matrices and the $4D$ volume element, respectively. The Dirac
action for a spin 1/2 fermion with a mass term can be expressed as
\cite{PRD_Oda_026002,Guo:2011qt}
\begin{eqnarray}
S_{\frac{1}{2}} = \int d^5 x \sqrt{-g} \left[\bar{\Psi} \Gamma^M
     \left(\partial_M+\omega_M\right) \Psi
     - M F(z) \bar{\Psi}\Psi\right]. \label{DiracAction}
\end{eqnarray}
Here $\omega_R=
\frac{1}{4} \omega_R^{\bar{R} \bar{S}} \Gamma_{\bar{R}}
\Gamma_{\bar{S}}$ is the spin connection, where 
\begin{eqnarray}
 \omega_R ^{\bar{R} \bar{S}}
  &=&
 - \frac{1}{2} {e}^{T \bar{R}} {e}^{Q \bar{S}}\partial_{[T} e_{Q]
{\bar{T}}} {e}_R^{~\bar{T}} + \frac{1}{2} {e}^{S \bar{[R}}\partial_{[R} e_{S]}^{~\bar{S]}}
\nonumber,
\end{eqnarray}
and $F(z)$ is some general scalar function, providing a mass term with a kink-like profile, which from this point is written in terms of a conformal variable $z$ such that $dz = \text{e}^{-A(y)} dy$ regards a transformation to conformal coordinates.
This kind of mass term is introduced in the action, for it has played a critical role
on the localization of fermionic fields on a Minkowski brane. The components of the spin connection
$\omega_M$ with respect to (\ref{metricaa}) are $ \omega_\mu =\frac{1}{2}(\partial_{z}A) \gamma_\mu \gamma_5
       +\hat{\omega}_\mu, $ where $\hat{\omega}_\alpha=\frac{1}{4} \bar\omega_\alpha^{\bar{\mu}
\bar{\nu}} \Gamma_{\bar{\mu}} \Gamma_{\bar{\nu}}$ is the spin
connection derived from the metric
${\rm g}_{\mu\nu}=\tilde{e}_{\mu}^{~\bar{\mu}}
\tilde{e}_{\nu}^{~\bar{\nu}}\eta_{\bar{\mu}\bar{\nu}}$. Thus, the
equation of motion corresponding to the action (\ref{DiracAction})
reads
\begin{eqnarray}
 \left[ \gamma^{\mu}(\partial_{\mu}+\hat{\omega}_\mu)
     + \gamma^5 \left(\partial_z +2 \partial_{z} A \right)
     -\text{e}^A MF(z)
 \right ] \Psi =0\,. \label{DEEE}
\end{eqnarray}

The $5D$ Dirac equation can be hence studied by taking spinors with respect to $4D$
effective fields. In this way the chiral splitting yields 
\begin{equation}
 \Psi= \text{e}^{-2A(z)}\left(\sum_n\psi_{Ln}(x^\mu) L_n(z)
 +\psi_{Rn}(x^\mu) R_n(z)\right),
\end{equation}
where $L_{n}(z)$ and $R_{n}(z)$ are the well-known KK modes, and $\psi_{Rn}(x^\mu)=+\gamma^5 \psi_{Rn}(x^\mu)$ [$\psi_{Ln}(x^\mu)=\gamma^5 \psi_{Ln}(x^\mu)$] is the 
right-chiral [left-chiral] component of a $4D$ Dirac field,
respectively. In addition, the sum over $n$ can be both continuous and discrete.
Assuming that $\gamma^\mu(\partial_\mu + \hat{\omega}_{\mu})\psi_{(R,L)n} = m_n\psi_{(L,R)n}$, the $L_n(z)$ and 
$R_n(z)$ functions should then satisfy the subsequent coupled equations, 
\begin{subequations}
\begin{eqnarray}
 \left[\partial_z
         - \text{e}^A MF(z) \right]R_n(z)
 &=& - m_n L_n(z). \label{bbb}\\ \left[\partial_z
         + \text{e}^A MF(z) \right]L_n(z)
 &=& ~~m_n R_n(z)\,. \label{aaa} 
\end{eqnarray}
\end{subequations}
The associated 
Schr\"{o}dinger-like equations can be thus acquired for the left and right-chiral KK
modes of fermions, respectively, as:
\begin{eqnarray}
 \big(-\partial^2_z + V_L(z) \big)L_n
      &=&m_{L_n}^{2} L_n\,,~~
  \label{left1} \\
 \big(-\partial^2_z + V_R(z) \big)R_n
      &=&m_{R_n}^{2} R_n\,,
  \label{fright}
\end{eqnarray}
where the mass-independent potentials are given by
\begin{subequations}\label{Vspin12}
\begin{eqnarray}
 V_L(z)&=& \text{e}^{2A} M^{2}F^{2}(z)
   - \text{e}^{A} A' M F(z) -\text{e}^{A}M\partial_{z}F(z), \label{VL}\\
 V_R(z)&=&  \text{e}^{2A} M^{2}F^{2}(z)
   + \text{e}^{A} A' M F(z) + \text{e}^{A}M\partial_{z}F(z). \label{VR}
\end{eqnarray}
\end{subequations}
Note that the Schr\"odinger-like equations \eqref{left1} and \eqref{fright} can be transformed into $U^\dagger U L_n = m_{n}^{2}L_n$ and $U U^\dagger R_n = m_{n}^{2}R_n$, where $U\equiv \partial_z
         + \text{e}^A MF(z)$. This observation is based upon supersymmetric quantum mechanics, implying that the 
mass squared is non-negative.

In order to lead these results to the standard $4D$ action for
a massless fermion, and a series of massive chiral fermions, the action $ S =\sum_{n}\int d^4 x \sqrt{-{\rm g}}
  ~\bar{\psi}_{n}
   \left[\gamma^{\mu}(\partial_{\mu}+\hat{\omega}_\mu)
    -m_{n}\right]\psi_{n}$ is employed, 
for orthonormalization conditions
\begin{eqnarray}
 \int_{-\infty}^{+\infty} L_m L_ndz
  &=& \delta_{mn} = 
 \int_{-\infty}^{+\infty} R_m R_ndz,\qquad 
 \int_{-\infty}^{+\infty} L_m R_ndz
  = 0. 
\end{eqnarray}

If in the formulae (\ref{bbb}) and (\ref{aaa}), by setting $m_n=0,$ thus it yields
\begin{eqnarray}
 L_0&\propto & {\rm e}^{-M \int {\rm e}^{A} F dz}, \label{zerol}\qquad R_0\propto {\rm e}^{M \int {\rm e}^{A} F dz}. \label{zeror}
\end{eqnarray}
Hence, either the massless left- or right-chiral
KK fermion modes can be localized on the brane, being the another one non-normalizable.

By taking $F(z)= \,\zeta(z)$, {regarding Eqs.~(\ref{015})-(\ref{015B}),} it yields 
\begin{subequations}\label{Vspin12}
\begin{eqnarray}
 V_L(z(y))
  &=& \text{e}^{2A(y)} \left(M^{2}\zeta^{2}(y) - \frac{dA(y)}{dy} M \zeta(y) - M\frac{d\zeta(y)}{dy}\right),
   \label{VL}\\
 V_R(z(y))&=& 
  \text{e}^{2A(y)} \left(M^{2}\zeta^{2}(y) + \frac{dA(y)}{dy} M \zeta(y) + M\frac{d\zeta(y)}{dy}\right). \label{VR}
\end{eqnarray}
\end{subequations}
Eqs. (\ref{VL}) and (\ref{VR}), evince that, when the mass
term in the action (\ref{DiracAction}) regards $M=0$, the
potentials for left and right-chiral KK modes $V_{L,R}(z)$ 
vanish. Then both chiral fermions cannot be
localized on the thick brane. Moreover, if $V_{L}(z)$ and
$V_{R}(z)$ are demanded to be $\mathbb{Z}_{2}$-even with respect to the extra dimension
$z$, then the mass term $MF(z)$ must be an odd function of $z$ {\cite{Liu:2013kxz}. In fact,  some useful classes of brane-world models have the extra dimension topology
$S^1/\mathbb{Z}_2$.  If the background scalar is an odd function of extra warped dimension, the  Yukawa coupling, between the fermion and the background scalar field, assures the  localization
mechanism for fermions \cite{Liu:2013kxz}.} {For the majority brane-world models, the scalar field $\zeta$  is, usually, a kink, being an odd function of the extra dimension. Here we do not necessarily impose this condition, in order to not preclude asymmetric solutions, with respect to the extra dimension.} 

\begin{figure}
\centerline{\psfig{file= 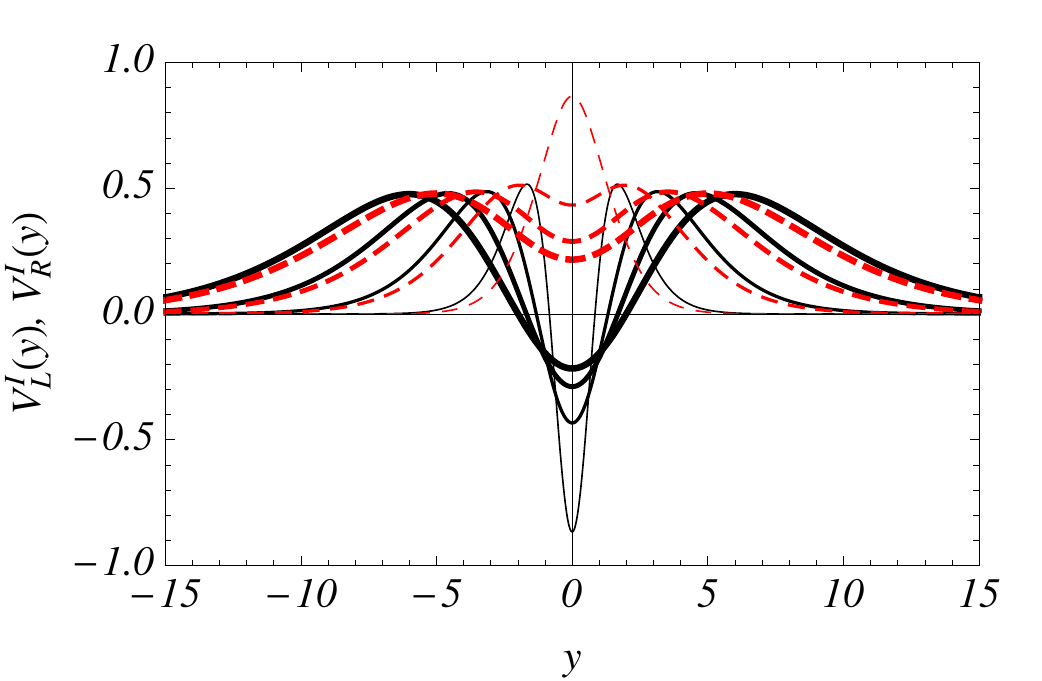, width= 8.5 cm}}
\centerline{\psfig{file= 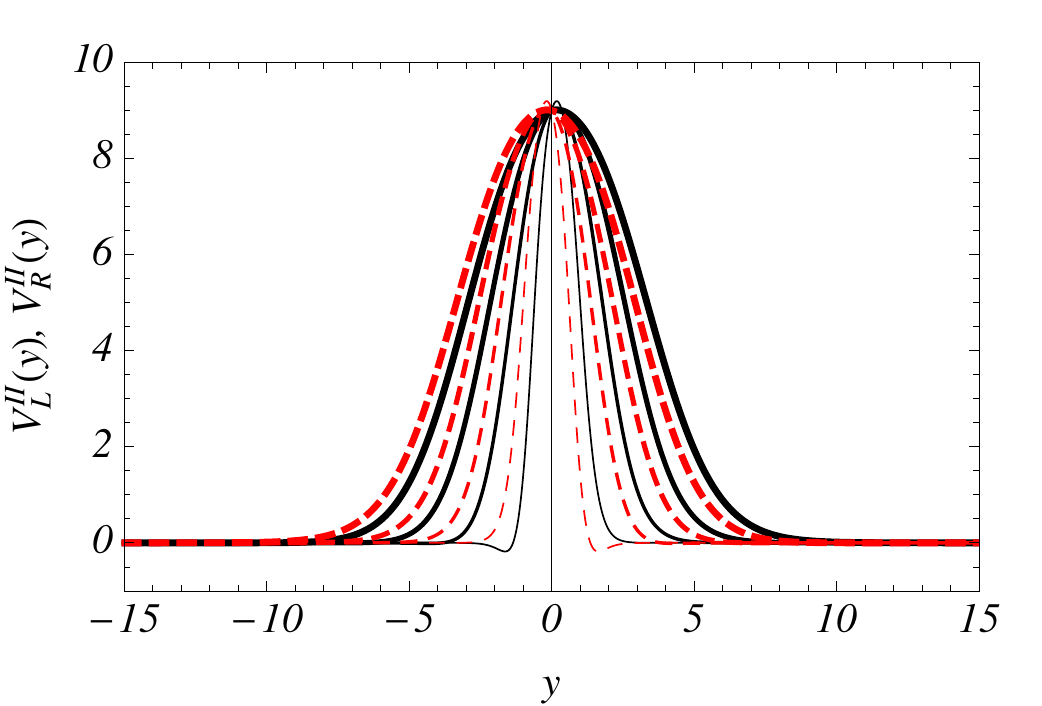, width= 8.3 cm}}
\centerline{\psfig{file= 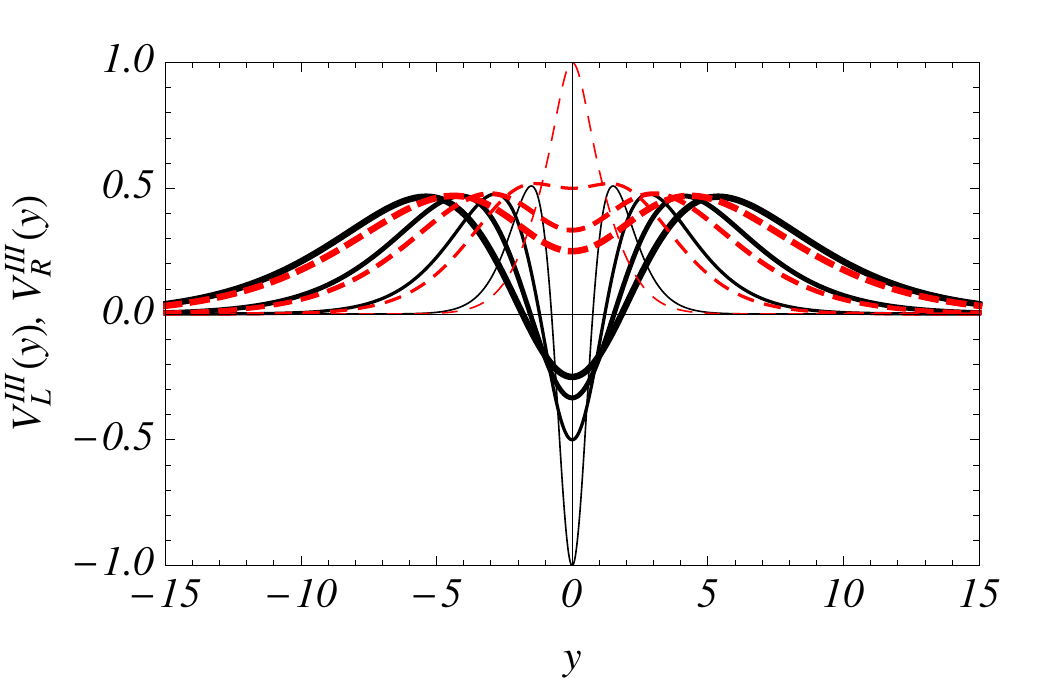, width= 8.5 cm}}
\centerline{\psfig{file= 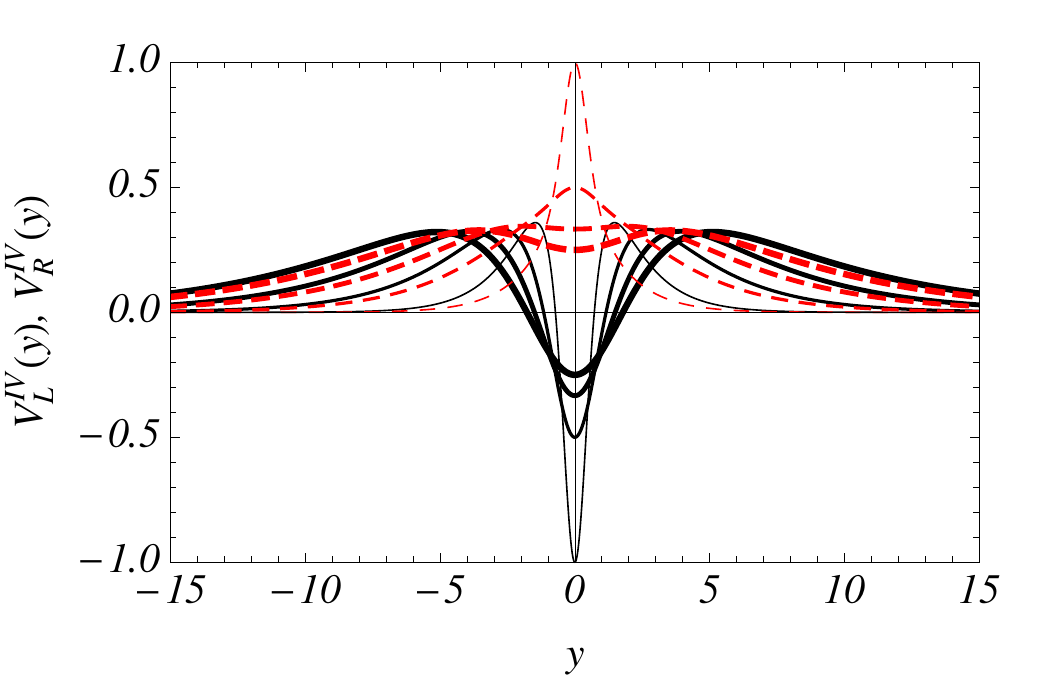, width= 8.5 cm}}
\caption{Associated Schr\"odinger-like quantum mechanical potential, $V_L(y)$ and $V_R(y)$, respectively for left-chiral (solid (black) lines) and right-chiral (dashed (red) lines) KK modes of fermions, for models from $I$ (first plot) to $IV$ (forth plot).
Again, one has considered integer values of the brane width parameter $a$, running from $1$ (thinest line) to $4$ (thickest line), corresponding to an increasing thickness.
The fermion mass parameter has been assumed to be equal to unit, $M=1$.}
\label{brane03}
\end{figure}

In what follows the profile of the above {\em left-right} potentials is depicted in Fig. (\ref{brane03}) for different values of the localization parameter $a$. In fact, the potentials $V_{L,R}(z)$ have asymptotic behaviors that tend to zero from up, as $y \to \pm\infty$, for all models from $I$ to $IV$. In the model $I$, at $y = 0$ the potential $V_R(z)$ attains its maximum positive value, a global maximum, for $a=1$. The potential $V_R(z)$ changes to a volcano-type profile along the interval of $1 < a < 2$, such that for $a = 2,\,3,\, 4,\dots$ the point $y=0$ regards a local minima, which allows for producing unstable resonances, which can be tunneled to the outside of the potential.
Nevertheless, the potential $V_L(z)$ has the associated minima at $y=0$ for all positive integer values of $a$, $a = 2,\,3,\, 4,\dots$, and it creates the conditions for producing bound states.
A very similar behavior is exhibited by the model $III$, in spite of showing different amplitudes. The model $IV$ is quite similar to these models, with the only qualitative difference concerning that, at $y = 0$, the potential $V_R(z)$ attains its maximum positive value, a global maximum, for $a=1$ and $a=2$.
For the model $IV$, the stability conditions created by the right- and left- chiral volcano-type potentials are more sensible to the increasing of the brane width ($a \gtrsim 3$), in comparison to models $I$ and $II$ ones ($a \gtrsim 2$), inducing no mass gap to separate the fermion zero mode from the excited KK massive modes. In these cases, there exist continuous spectra for the Kaluza-Klein modes of fermions of both chiralities. These volcano-type potentials imply into the existence of resonant or metastable states of fermions which can tunnel from the brane to
the bulk \cite{Liu_20101}. 
The left-chiral KK mode has a continuous gapless spectrum for the models $I$, $III$, and $IV$, according to Fig. (\ref{brane03}). Since the potential for left-chiral fermions presents a negative value at the brane location for these models, the zero mode of right- and left-chiral fermions, $R_0(y)$ and $L_0(y)$ are the only necessary ingredient to be tested to be localized on the brane. 
For the model $II$, both potentials for the left- and right-chiral fermions have positive values of the potential, irrespectively of $y$.
However for, in both cases, i. e. for $V_{R,L}(y)$, when $a \lesssim 1$, an asymmetric behavior emerges and produces a totally odd symmetric well-barrier profile in the limit of $a \to 0$. 
Except for $0 < a \lesssim 1$, the zero mode of left- and right-chiral fermions can not be trapped. 
All potentials for the model $II$ are asymmetric (except for $a=0$, which is nonsense in the brane context), have maxima at $y=0$ and tends to zero at $y\to \pm \infty$, and there is no bound state for right-chiral fermions.
In particular, for $V_{R,L}(y)$ when $a=1$, the minima occur at $y\sim \pm 0.87$.

\section{Matter localization for spin 0 scalar fields}

The localization of scalar fields on thick branes generated by deformed defects can also be considered from this point. In particular, an interesting approach on domain walls can be also found in Ref. \cite{PRD_Stojkovic}.
In fact, a massive scalar field coupled to gravity can be described by the following action,
\begin{eqnarray}\label{escalar_lagrangiana}
S_{0}=-\frac{1}{2}\int d^{5}x\sqrt{-g}~
     \left(g_{MN}\partial^{M}\Phi\partial^{N}\Phi+m_0 ^2\Phi^2\right),
\end{eqnarray}
where $m_0$ denotes the effective mass of a bulk scalar field, $\Phi$, and from where one can check whether spin 0 matter fields can be trapped on the thick brane.
By employing the metric (\ref{metricaa}), the associated 
equation of motion from the action in Eq.~(\ref{escalar_lagrangiana}) reads
\begin{eqnarray}\label{EqOfScalar5D}
\frac{1}{\sqrt{-{\rm g}}}\partial_{\mu}\left(\sqrt{-{\rm g}} {\rm g}^{\mu \nu}\partial_{\nu}
\Phi\right) + {\rm e}^{-3A}\partial_{z} \left({\rm e}^{3A}\partial_z\Phi\right) - {\rm e}^{2A}m_0 ^2\Phi = 0.
\end{eqnarray}
Hence, by the KK decomposition
 $\Phi(x^\mu,z)=\sum_{n}\chi_{n}(x^\mu)\xi_{n}(z){\rm e}^{-3A/2}$, where 
 $\xi_{n}$ is assumed to satisfy the $4D$ Klein-Gordon
equation $
\left[\partial_{\mu}\left(\sqrt{-{\rm g}}
  {\rm g}^{\mu \nu}\partial_{\nu}\right)/{\sqrt{-{\rm g}}} -m_{n}^{2} \right]\xi_{n}(x^\mu)=0,$ being $m_{n}$ the $4D$ mass of the KK excitation of the scalar field. Then the scalar KK mode $\xi_{n}(z)$ is ruled by the following equation: 
\begin{eqnarray}
\left[-\partial^{2}_z+ V_{0}(z)\right]{\xi}_n(z)
 =m_{n}^{2} {\xi}_{n}(z)\,.
 \label{SchEqScalar1}
\end{eqnarray}
This equation is a Schr\"{o}dinger one, with effective potential given by 
\begin{eqnarray}
 V_0(z(y))&=& \frac{3}{2} A''(z) + \frac{9}{4}A'^{2}(z) + {\rm e}^{2A(z)}m_0 ^2= {\rm e}^{2A(z)}\left(\frac{3}{2}\frac{d^2A(y)}{dy^2} + \frac{15}{4}\left(\frac{dA(y)}{dy}\right)^{2} + m_0 ^2\right). \label{potencialv0}
\end{eqnarray}

\begin{figure}
\centerline{\psfig{file= 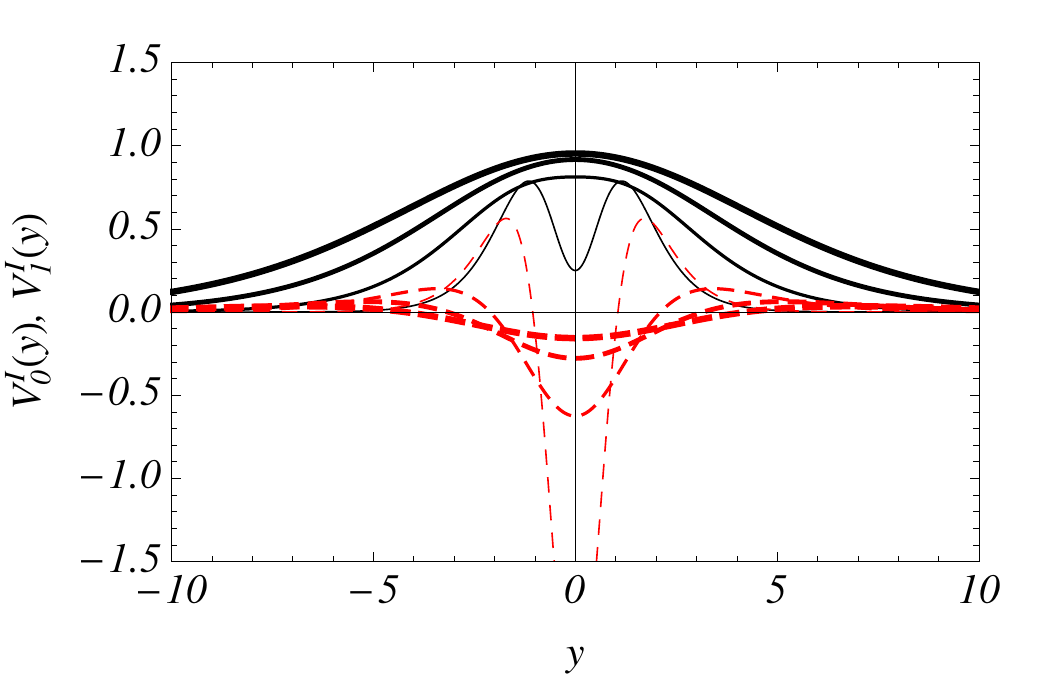, width= 8.5 cm}}
\centerline{\psfig{file= 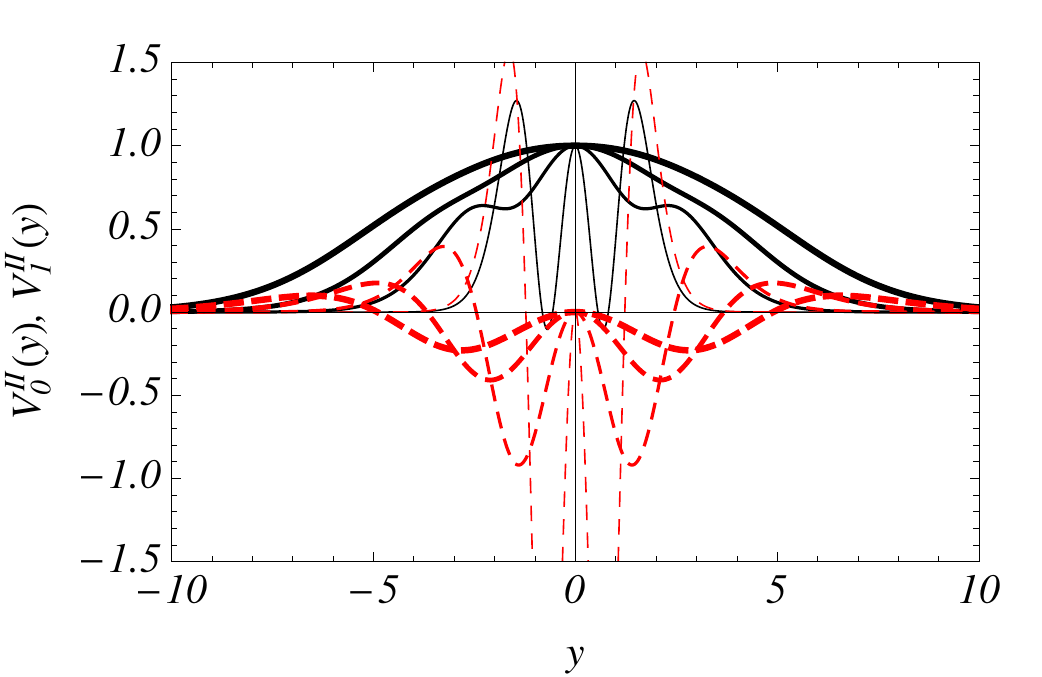, width= 8.5 cm}}
\centerline{\psfig{file= 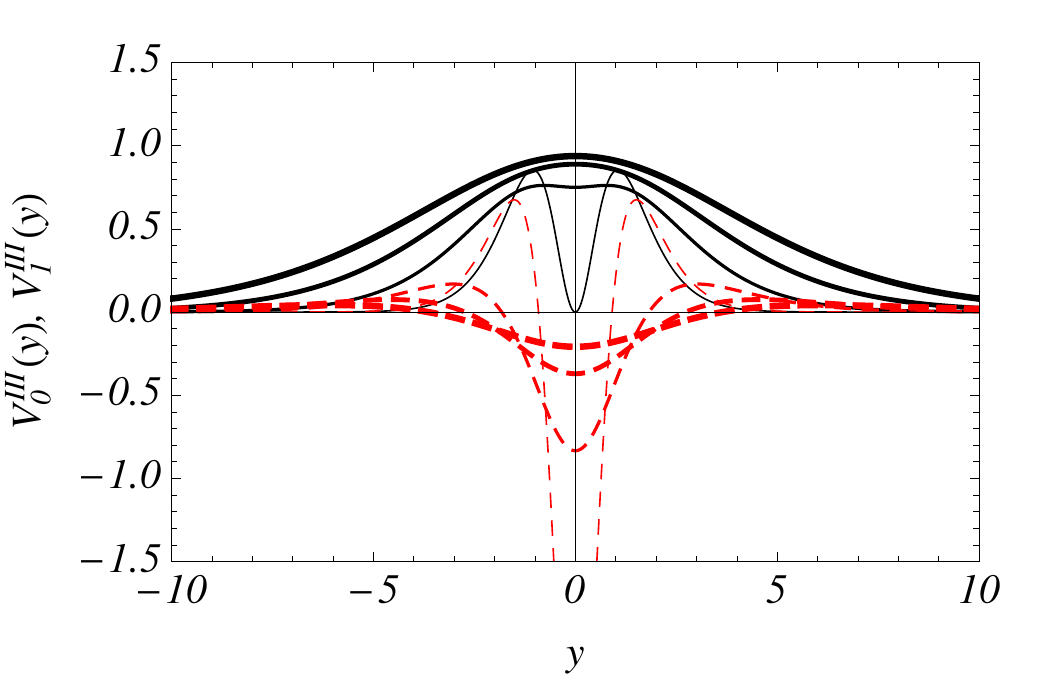, width= 8.5 cm}}
\centerline{\psfig{file= 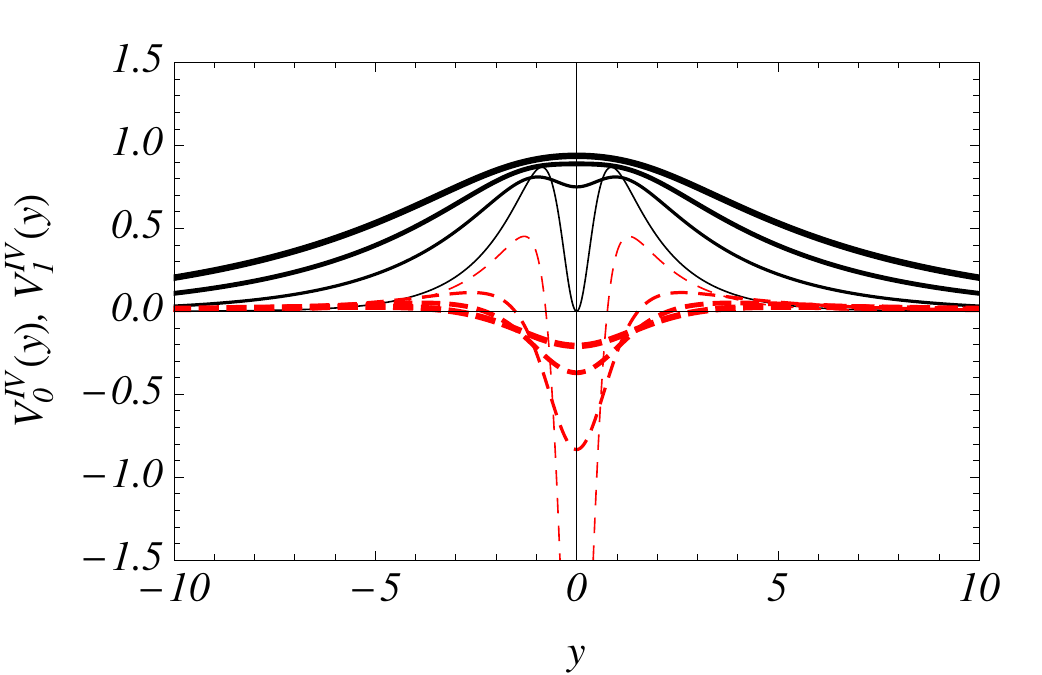, width= 8.5 cm}}
\caption{Associated Schr\"odinger-like quantum mechanical potential, $V_0(y)$ and $V_0(y)$, respectively for spin 0 (solid (black) lines) and spin 1 (dashed (red) lines) bosons for models from $I$ (first plot) to $IV$ (forth plot).
Again, one has considered integer values of the brane width parameter $a$ running from $1$ (thinest line) to $4$ (thickest line), corresponding to an increasing thickness.
The spin 0 boson mass parameter has been assumed to be equal to unit.}
\label{brane04}
\end{figure}

The profile of the above scalar boson potential is depicted in Fig. (\ref{brane04}) (solid (black) lines) for different values of the localization parameter $a$.
For $m_0 = 1$, only brane scenarios with $a \lesssim 2$ provide conditions to have a localized scalar field.
Even in this case, such localized states behave much more as resonances than as bound states, given that it can be tunneled out of the potential.
Bound states appear only for non integer values of the brane width such that $a < 1$, which shall correspond to typical volcano-type potentials.

\section{Matter localization for spin 1 vector fields}

One now turns to spin 1 vector fields and begins with the
$5D$ action of a vector field
\begin{eqnarray}\label{spin1_Lag}
S_{1} = -\frac{1}{4}\int d^{5}x \sqrt{-g}~ g^{M N}
 g^{RS}F_{MR}F_{NS},
\end{eqnarray}
where $F_{MN}=\partial_{[M}A_{N]}$ denotes the field
strength tensor. 
A $5D$ spin 1 field can be now studied via the KK decomposition $
 A_{M}(x^{\rho},z)=\sum_{n} a_{M}^{(n)}(x^{\rho})\tau_n (z).$ 
The action of the $5D$ massless vector field (\ref{spin1_Lag}) is invariant under the following
gauge transformation:
\begin{eqnarray}
 A_{M}(x^{\rho},z) \mapsto \tilde{A}_{M}(x^{\rho},z) &=&
 A_{M}(x^{\rho},z)+\partial_{M}F(x^{\rho},z), \label{gauge5}
\end{eqnarray} where $F(x^{\rho},z)$ denotes any arbitrary regular scalar function, for $M=\mu,5$. The field component $A_{5}(x^{\rho},z)$ equals zero \cite{Guo:2011qt}, by this gauge. In fact, Eq.~(\ref{gauge5}) yields 
\begin{eqnarray}
 \tilde{A}_{5}(x^{\rho},z)
  = \sum_{n}a_{5}^{(n)}(x^{\rho})\tau_n (z)+\partial_{z}F(x^{\rho},z).
\end{eqnarray}
By choosing 
 $F(x^{\rho},z)= -\sum_{n}
      a_{5}^{(n)}(x^{\rho})\int\tau_n (z) dz, \label{F(x,z)}$ \cite{Guo:2011qt} 
then $\tilde{A}_5=0$, hence the action (\ref{spin1_Lag}) is led to the spacetime action
\begin{eqnarray}
S_1 = - \frac{1}{4} \int d^5 x \sqrt{-g} \left(
    F^{\mu\nu}F_{\mu\nu}
    +2{\rm e}^{-A} g^{\mu\nu} A'_{\mu} A'_{\nu}
    \right).
\label{acaospin1_2}
\end{eqnarray} where $(\;\;)' = \partial_z$. 
Given a set of orthonormal functions $\tau_n (z)$, playing the role of spin 1 Kaluza-Klein modes, and the decomposition of the vector field $A_{\mu}(x^\rho,z)=\sum_n
a^{(n)}_\mu(x^\mu)\tau_n (z){\rm e}^{-A/2}$, 
the action (\ref{acaospin1_2}) reads
\begin{eqnarray}
S_1 = -\frac12\sum_{n}\int d^4 x \sqrt{-{\rm g}}~
    \bigg(\frac{1}{2}
       {f^{(n)}}^{\mu\nu}f^{(n)}_{\mu\nu}
    +m_{n}^2 {a^{(n)}}^{\mu}a^{(n)}_{\mu}
    \bigg),
\nonumber
\end{eqnarray}
where $f^{(n)}_{\mu\nu} = \partial_{[\mu} a^{(n)}_{\nu]}$ stands for the $4D$ field strength tensor. The KK modes
$\tau_n (z)$ satisfy the Schr\"odinger equation
\begin{eqnarray}
  \left(-\partial^2_z +V_1(z) \right){\tau}_n(z)=m_n^{2}  {\tau}_n(z),
  \nonumber
\end{eqnarray}
where the mass-independent potential reads {{\cite{guo}
\begin{eqnarray}V_1(z)&=& \frac{1}{4}A'^2(z) + \frac{1}{2}A''(z)= {\rm e}^{2A(z)}\left[\frac{1}{2}\frac{d^2A(y)}{dy^2} + \frac{3}{4}\left(\frac{dA(y)}{dy}\right)^{2} \right].
 \end{eqnarray}
}

The profile of the above vector boson potential is depicted in Fig. (\ref{brane04}) (dashed (red) lines) for different values of the localization parameter $a$.
All the thick brane scenarios with $a \gtrsim 1$ provide localization conditions to have vector field bound states.
Increasing values of $a$ lead to more stable bound states.}

\section{Conclusions and discussion}
\label{SecConclusion}

Thick branes driven by superpotentials supported by deformed defects (c. f. Eqs. (\ref{014})-(\ref{sp002})), for various bulk matter fields of spin 0, spin 1/2, and spin 1 have been investigated. For spin 1 gauge fields, the profile of the associated vector boson potential showed that, in the thick brane models for $a\gtrsim 1$, localization conditions hold as to guarantee the existence of vector field bound states. Quantitatively, increasing values of the brane thickness parameter, $a$, leads to more stable bound states.

Concerning spin 0 (scalar) fields, the profile of the potential evinces that only thick brane scenarios with $a\gtrsim 4$ provide localization conditions compatible to scalar field bound states. 

The most intricate result is related to spin 1/2 fields. In fact, for fermionic fields, {left-right} potentials were deeply studied and for the four models considered here, the issue of localization has been scrutinized. It is worth to point out that models $I$, $III$, and $IV$ admit volcano-type potentials, inducing no mass gap to separate the fermion zero mode from the excited KK massive modes. Hence a continuous spectra
for the Kaluza-Klein modes of fermions of both chiralities are allowed. A refined analysis of the values of the $a$ parameter in these models, influencing the localization of fermionic fields, was provided. The model $II$, induced by the superpotential (\ref{0141}), reveals a peculiar behavior. In this model, right-chiral fermions have positive values of the potential irrespectively of the extra dimension when $a=1$. Hence, except for this value, the zero mode of left- and right-chiral fermions can not be trapped. All potentials for the model $II$ are asymmetric, have maxima at $y=0$ and minima at $y\to \pm \infty$, and there is no bound state for right-chiral fermions, but, again, for $V_{R,L}(y)$ when $a=1$.

It is worth to mention that, for the localization of a fermion zero mode, the  mass term $MF(z)\bar\Psi\Psi$ was considered in the $5D$ action. 
An interesting approach concerning such mass term in Eq.~(\ref{DiracAction}) has been studied, corresponding to the so-called singular dark spinors \cite{Ahluwalia:2009rh,daRocha:2011yr}. Such massive mass dimension one quantum fields are prime candidates for the dark matter problem, also presenting possible signatures at LHC \cite{Dias:2010aa}. It generates a slightly different action responsible for spin 1/2 matter fields localization \cite{Liu:2011nb,Jardim:2014xla}. Such approach can be also extended in the context of deformed defects here presented.
 
\section*{Acknowledgement}
{\em Acknowledgments - The work of AEB is supported by the Brazilian Agencies FAPESP (grant 2015/05903-4) and CNPq (grant No. 300809/2013-1 and grant No. 440446/2014-7). RdR is grateful to CNPq (grants No. 303293/2015-2, and No. 473326/2013-2), and to FAPESP (grant 2015/10270-0) for partial financial support.}


\end{document}